# Iterative Mode-Dropping for the Sum Capacity of MIMO-MAC with Per-Antenna Power Constraint

Yang Zhu and Mai Vu

Department of Electrical and Computer Engineering, McGill University, Montréal, Canada, H3A 2A7

E-mail: yang.zhu@mail.mcgill.ca, mai.h.vu@mcgill.ca

*Abstract*—We propose an iterative mode-dropping algorithm that optimizes input signals to achieve the sum capacity of the MIMO-MAC with per-antenna power constraint. The algorithm successively optimizes each user's input covariance matrix by applying mode-dropping to the equivalent single-user MIMO rate maximization problem. Both analysis and simulation show fast convergence. We then use the algorithm to briefly highlight the difference in MIMO-MAC capacities under sum and per-antenna power constraints.

## I. INTRODUCTION

Under sum power constraint, the capacity of MIMO channels has already been well-established. With channel state information (CSI) available at both the transmitter and the receiver, the single-user capacity is achieved by diagonalizing the channel matrix then water-filling the total power across the spatial subchannels [1]. Based on this single-user water-filling algorithm, the sum capacity of the MIMO multiple access channel (MIMO-MAC) can be efficiently computed in an iterative manner, by optimizing the transmit covariance matrix of each user successively while treating the signals from all other users as noise [2].

Sum power constraint is, however, less practical than per-antenna constraint because the power amplifier at each antenna has its own output dynamic range [3]. The capacity under this per-antenna power constraint has been investigated for different channel models. For single-user MISO channels, the capacity can be expressed in closed form and is achieved by a single-mode beamformer, with its phase matched to that of the channel vector and its amplitude determined by the power constraint [4]. For single-user MIMO channels, the capacity optimization problem can be formulated in the framework of semidefinite programming and a simple adaptive *mode-dropping* algorithm computes the capacity by solving the Karush-Kuhn-Tucker (KKT) conditions efficiently [5]. For the MIMO-MAC, although the sum-rate maximization problem is convex and hence readily solvable using interior-point methods, the computational complexity could be impractically high when the system has a large number of users or antennas.

In this regard, we propose a more efficient algorithm called *iterative mode-dropping* to find the optimal transmit covariance matrices that achieve the sum capacity of the MIMO-MAC with per-antenna power constraint. In a way similar to the iterative water-filling algorithm [2], iterative mode-dropping optimizes the transmit covariance matrices successively. In each step, it optimizes the covariance matrix of one user based on the mode-dropping algorithm for single-user MIMO channels, while holding those of the other users constant. The algorithm converges surely to the sum capacity and simulation results show fast convergence, typically within $10^{-6}$ bps/Hz of the sum capacity after 10 iterations (i.e., each user has updated its covariance matrix 10 times). We also use the algorithm to compare the sum capacity under different power constraints and study the effects on capacity of SNR, the number of users and the number of antennas.

The following notation is used: an italic letter, boldface lowercase or boldface uppercase letter represents a scalar, column vector or matrix, respectively. $(\cdot)^T$ or $(\cdot)^\dagger$ denotes transpose or conjugate transpose of a matrix. $\text{Tr}(\cdot)$ is the trace and $\det(\cdot)$ is the determinant of a matrix. The operator $\succcurlyeq$ and $\preccurlyeq$ represent positive semidefinite ordering of Hermitian matrices. $\text{diag}\{\cdot\}$ or $\text{diag}(\cdot)$ forms a diagonal matrix with given entries or from the diagonal entries of the given matrix, respectively. $\mathbf{I}$ is an identity matrix of appropriate dimension.

## II. SYSTEM MODEL AND KNOWN CAPACITY

### A. MIMO-MAC model and per-antenna power constraint

A $K$-user Gaussian MIMO-MAC with $n$ antennas at each transmitter and $m$ antennas at the receiver can be modeled as

$$\mathbf{y} = \sum_{i=1}^{K} \mathbf{H}_i \mathbf{x}_i + \mathbf{z}, \qquad (1)$$

where vector $\mathbf{x}_i \in \mathbb{C}^{n \times 1}$ is the transmitted signal of the $i$th user, vector $\mathbf{y} \in \mathbb{C}^{m \times 1}$ is the received signal, $\mathbf{H}_i \in \mathbb{C}^{m \times n}$ is the channel matrix from the $i$th user to the receiver, and vector $\mathbf{z} \sim \mathcal{CN}(\mathbf{0}, \mathbf{I})$ is the additive circularly symmetric complex Gaussian noise at the receiver[1]. We assume that the channel matrices are full-rank and perfect CSI is available at all transmitters and the receiver.

In this model, we consider a per-antenna power constraint in which each transmit antenna has its own power budget and the antennas of the same user can cooperate in terms of signaling. Considering zero-mean input signals, $\mathbf{Q}_i = E[\mathbf{x}_i \mathbf{x}_i^\dagger]$ is the transmit covariance matrix of the $i$th user. Hence, the set of covariance matrices $\{\mathbf{Q}_i\}$ satisfies

$$\text{diag}(\mathbf{Q}_i) \preccurlyeq \mathbf{P}_i, \quad \forall i = 1, \cdots, K, \qquad (2)$$

---
[1]Without loss of generality, we assume that this noise vector is spatially white and has been normalized.

where $\mathbf{P}_i = \text{diag}\{P_{i1}, P_{i2}, \cdots, P_{in}\}$ and $P_{ij}$ is the power constraint for the $j$th antenna of the $i$th user.

### B. Review of the capacity with sum power constraint

For comparison, the sum power constraint and the corresponding MIMO-MAC capacity are summarized here. In this scenario, the $i$th user has the total transmit power across all its antennas constrained by $P_i$, that is,

$$\text{tr}(\mathbf{Q}_i) \leq P_i, \quad i = 1, \cdots, K . \quad (3)$$

To achieve sum capacity, the users choose their positive semidefinite transmit covariance matrices, $\{\mathbf{Q}_i\}$, to maximize the sum rate of the MIMO-MAC subject to the above power constraints.

This problem can be solved efficiently by iterative water-filling [2]. The underlying principle is to successively update each one of the $K$ covariance matrices, say $\mathbf{Q}_k$, by keeping all other $K-1$ matrices at the other users unchanged. In each step, the subproblem is equivalent to a single-user rate maximization problem by treating other users' signals as noise, for which water-filling is readily applicable. This algorithm converges surely to the sum capacity and the rate of convergence is fast.

## III. SUM RATE MAXIMIZATION WITH PER-ANTENNA POWER CONSTRAINT

The sum-rate maximization problem for MIMO-MAC with per-antenna power constraint can be formulated as a convex optimization problem of $\{\mathbf{Q}_i\}$:

$$\max_{\{\mathbf{Q}_i\}} \quad \log\det\left(\mathbf{I}_m + \sum_{i=1}^{K} \mathbf{H}_i \mathbf{Q}_i \mathbf{H}_i^\dagger\right) \quad (4)$$

$$\text{s.t.} \quad \text{diag}(\mathbf{Q}_i) \preccurlyeq \mathbf{P}_i, \quad i = 1, \cdots, K$$

$$\mathbf{Q}_i \succcurlyeq \mathbf{0}, \quad i = 1, \cdots, K .$$

The sum capacity with per-antenna power constraint is usually less than that with sum power since the feasible region of the former optimization problem is included in the latter as a subset. Unfortunately, no closed-form solution has been found for this problem. Inspired by iterative water-filling, we propose an iterative mode-dropping algorithm to solve this problem efficiently. Before presenting this algorithm, it is necessary to briefly review mode-dropping for single-user MIMO channels as in [5].

### A. Single-user mode-dropping

The problem in (4) reduces to the scenario of single-user MIMO channels when $K = 1$. The KKT conditions are necessary and sufficient for optimality because Slater's condition is satisfied, and the objective function and inequality constraints are continuously differentiable and concave functions of $\{\mathbf{Q}_i\}$. Based on the KKT conditions, a set of single-user optimality conditions can be stated as follows.

$$\mathbf{H}_1^\dagger (\mathbf{I}_m + \mathbf{H}_1 \mathbf{Q}_1^* \mathbf{H}_1^\dagger)^{-1} \mathbf{H}_1 = \mathbf{D}_1 - \mathbf{M}_1 \quad (5)$$

$$\mathbf{M}_1 \mathbf{Q}_1^* = \mathbf{0}$$

$$\text{diag}(\mathbf{Q}_1^*) = \mathbf{P}_1$$

$$\mathbf{M}_1, \mathbf{Q}_1^* \succcurlyeq \mathbf{0}$$

$$\text{diagonal } \mathbf{D}_1 \succ \mathbf{0} ,$$

where positive-definite diagonal matrix $\mathbf{D}_1 \in \mathbb{C}^{n \times n}$ is the dual variable associated with the per-antenna power constraint and Hermitian positive semidefinite matrix $\mathbf{M}_1 \in \mathbb{C}^{n \times n}$ is the dual variable associated with the positive semidefinite constraint. The covariance matrix $\mathbf{Q}_1^*$ satisfying the optimality conditions in (5) is the optimal solution that achieves the single-user MIMO capacity.

The most important and challenging step of mode-dropping is to express $\mathbf{Q}_1^*$ as a function of $\mathbf{D}_1$ by eliminating $\mathbf{M}_1$ in the optimality conditions. In [5], closed-form solution for $\mathbf{Q}_1^*$ in terms of $\mathbf{D}_1$ is established for all channel sizes. Using this solution, the remaining step is to find the optimal dual variable $\mathbf{D}_1$ such that the established solution satisfies the power constraint. To arrive at this optimal value, an iterative algorithm is proposed to update $\mathbf{D}_1$ at each step until the duality gap converges to within a specified tolerance. Due to space limitation, we summarize the main steps of this algorithm as follows:

1) *Compute $\mathbf{Q}_1^*$ from $\mathbf{D}_1$:* Let the singular value decomposition (SVD) of the channel matrix be $\mathbf{H}_1 = \mathbf{U}_\mathbf{H} \mathbf{\Sigma}_\mathbf{H} \mathbf{V}_\mathbf{H}^\dagger$. If $m \geq n$, let $\mathbf{\Sigma}_\mathbf{H} = [\mathbf{\Sigma}_n, \mathbf{0}_{n,m-n}]^T$ and form
   - $\mathbf{K} = \mathbf{V}_\mathbf{H} \mathbf{\Sigma}_n \mathbf{V}_\mathbf{H}^\dagger$, $\mathbf{K}^{-1} = \mathbf{V}_\mathbf{H} \mathbf{\Sigma}_n^{-1} \mathbf{V}_\mathbf{H}^\dagger$.
   - $\mathbf{F}_n = \mathbf{K} \mathbf{D}_1^{-1} \mathbf{K}^\dagger$.
   - $-\mathbf{S}_n =$ non-positive eigenmodes of $(\mathbf{F}_n - \mathbf{I}_n)$, here the modes contained in $\mathbf{S}_n$ are dropped.
   - $\mathbf{Z} = \mathbf{K}^{-1} \mathbf{S}_n (\mathbf{K}^{-1})^\dagger$.
   - $\mathbf{Q}_1^* = \mathbf{D}_1^{-1} - \mathbf{K}^{-1} (\mathbf{K}^{-1})^\dagger + \mathbf{Z}$.

   If $m < n$, let $\mathbf{\Sigma}_\mathbf{H} = [\mathbf{\Sigma}_m, \mathbf{0}_{m,n-m}]$ and $\mathbf{V}_1$ be the first $m$ columns of $\mathbf{V}_\mathbf{H}$ (i.e., write $\mathbf{V}_\mathbf{H} = [\mathbf{V}_1 \ \mathbf{V}_2]$).
   - Define the pseudo inverse of $\mathbf{H}_1$ as $\mathbf{H}^{-1} = \mathbf{V}_1 \mathbf{\Sigma}^{-1}{}_m \mathbf{U}_\mathbf{H}^\dagger$.
   - Take $(-\mathbf{S}_m)$ as the non-positive eigenmodes of $\mathbf{H}_1 \mathbf{D}_1^{-1} \mathbf{H}_1^\dagger - \mathbf{I}_m$.
   - $\mathbf{Z} = \mathbf{H}^{-1} \mathbf{S}_m (\mathbf{H}^{-1})^\dagger$.
   - $\mathbf{B} = \mathbf{V}_1^\dagger \left(\mathbf{Z} - \mathbf{H}^{-1}(\mathbf{H}^{-1})^\dagger\right) \mathbf{D}_1 \mathbf{V}_2 (\mathbf{V}_2^\dagger \mathbf{D}_1 \mathbf{V}_2)^{-1}$.
   - $\mathbf{A} = \left(\mathbf{I}_{n-m} - \mathbf{B}^\dagger \mathbf{V}_1^\dagger \mathbf{D}_1 \mathbf{V}_2\right) (\mathbf{V}_2^\dagger \mathbf{D}_1 \mathbf{V}_2)^{-1}$.
   - $\mathbf{X} = \mathbf{V}_2 \mathbf{A} \mathbf{V}_2^\dagger + \mathbf{V}_1 \mathbf{B} \mathbf{V}_2^\dagger + \mathbf{V}_2 \mathbf{B} \mathbf{V}_1^\dagger$.
   - $\mathbf{Q}_1^* = \mathbf{D}_1^{-1} - \mathbf{H}^{-1}(\mathbf{H}^{-1})^\dagger + \mathbf{Z} - \mathbf{X}$.

2) *Update $\mathbf{D}_1$:* $\mathbf{D}_1^{-1} \leftarrow \mathbf{D}_1^{-1} + \mathbf{P}_1 - \text{diag}(\mathbf{Q}_1^*)$.

After choosing an initial $\mathbf{D}_1 \succ \mathbf{0}$, the above procedure is repeated until the duality gap $|\text{tr}(\mathbf{D}_1(\mathbf{Q}_1^* - \mathbf{P}_1))|$ converges to within a pre-specified tolerance. The algorithm is shown to always converge to the optimal value. For more details on the derivation and convergence analysis, refer to [5].

The capacity of single-user MISO channels with per-antenna constraint is a special case of $n > m$ in which $m = 1$ and the channel matrix $\mathbf{H}_1$ reduces to a vector $\mathbf{h}_1 \in \mathbf{C}^{1 \times n}$. In this case, closed-form optimal solution exists and no iterative procedure is necessary [4]. The entries of optimal covariance matrix are given by

$$q_{ij} = \frac{h_{1i}^* h_{1j}}{|h_{1i} h_{1j}|} \sqrt{P_{1i} P_{1j}}, \quad i,j = 1, \cdots, n. \quad (6)$$

where the $h_{1i}$ is the $i$th entry of channel vector $\mathbf{h}_1$.

## B. Iterative mode-dropping

The following theorem on the optimality conditions motivates and lays the foundation for the iterative mode-dropping algorithm:

**Theorem 1** (Optimality conditions). *The set of transmit covariance matrices, $\{\mathbf{Q}_i^*\}$, is the optimal solution of the MIMO-MAC sum-rate maximization problem in (4) if and only if for each $i \in \{1, \cdots, K\}$, $\mathbf{Q}_i^*$ is the optimal solution of the single-user rate maximization problem:*

$$\begin{aligned}
\max \quad & \log\det(\mathbf{I}_m + \hat{\mathbf{H}}_i \mathbf{Q}_i \hat{\mathbf{H}}_i^\dagger) \quad (7)\\
\text{s.t.} \quad & \operatorname{diag}(\mathbf{Q}_i) \preccurlyeq \mathbf{P}_i \\
& \mathbf{Q}_i \succcurlyeq \mathbf{0} \; ,
\end{aligned}$$

*where $\hat{\mathbf{H}}_i$ is the effective channel matrix defined as*

$$\hat{\mathbf{H}}_i = \Big(\mathbf{I}_m + \sum_{k=1, k \neq i}^{K} \mathbf{H}_k \mathbf{Q}_k^* \mathbf{H}_k^\dagger\Big)^{-1/2} \mathbf{H}_i \; . \quad (8)$$

*Proof:* The *only if* part is easy and follows similar argument to the proof of Theorem 1 in [2]. Herein, we prove the *if* part. From the KKT conditions of the multiuser problem in (4), we can obtain a set of necessary and sufficient optimality conditions as

$$\mathbf{H}_i^\dagger \Big(\mathbf{I}_m + \sum_{i=1}^{K} \mathbf{H}_i \mathbf{Q}_i^* \mathbf{H}_i^\dagger\Big)^{-1} \mathbf{H}_i = \mathbf{D}_i - \mathbf{M}_i \quad (9)$$

$$\mathbf{M}_i \mathbf{Q}_i^* = \mathbf{0}$$

$$\operatorname{diag}(\mathbf{Q}_i^*) = \mathbf{P}_i$$

$$\mathbf{M}_i, \mathbf{Q}_i^* \succcurlyeq \mathbf{0}$$

$$\text{diagonal } \mathbf{D}_i \succ \mathbf{0} \; ,$$

for all $i = 1, 2 \cdots, K$. The matrices $\{\mathbf{M}_i\}$ and $\{\mathbf{D}_i\}$ are dual variables associated with the per-antenna power constraints of the $K$ users and the positive semidefinite constraints of the covariance matrices. Using similar arguments to the single-user case, a set of covariance matrices satisfying these conditions is the optimal solution of the MIMO-MAC problem.

Next, we decouple the conditions in (9) into an equivalent ensemble of $K$ single-user optimality conditions. This can be done by rewriting the first equation in (9) as

$$\hat{\mathbf{H}}_i^\dagger (\mathbf{I_m} + \hat{\mathbf{H}}_i \mathbf{Q}_i^* \hat{\mathbf{H}}_i^\dagger)^{-1} \hat{\mathbf{H}}_i = \mathbf{D}_i - \mathbf{M}_i \; , \quad (10)$$

where $\hat{\mathbf{H}}_i$ is the effective channel matrix defined in (8). It follows that $\operatorname{rank}(\hat{\mathbf{H}}_i) = \operatorname{rank}(\mathbf{H}_i)$. By replacing the first equation in (9) by (10), the multiuser optimality conditions are successfully decoupled.

In turn, the new conditions with (10) for each $i$ are exactly the optimality conditions for the single-user rate maximization problems in (7). That is, if for each $i \in \{1, \cdots, K\}$, $\mathbf{Q}_i^*$ is the optimal solution of (7), then the set $\{\mathbf{Q}_i^*\}$ will automatically satisfy the multiuser optimality conditions in (9), and is therefore the optimal solution of the sum-rate problem in (4). ∎

Theorem 1 uncovers a simple but interesting fact: the optimal transmit covariance matrices, $\{\mathbf{Q}_i^*\}$, are in a state of equilibrium. The process of finding them can be viewed as that of moving towards this equilibrium state. The iterative mode-dropping algorithm successively optimizes one of these matrices at a time, while maintaining the others, and in the end reaches the equilibrium. The procedure is given in Algorithm 1.

---

**Algorithm 1:** Iterative mode-dropping for MIMO-MAC sum-rate maximization with per-antenna power constraint

Initialize $\mathbf{Q}_i = \mathbf{0}$, $i = 1, \cdots, K$
**repeat**
  for $i = 1, \cdots, K$
    $\mathbf{W}_z^{(i)} = \mathbf{I}_m + \sum_{k=1, k\neq i}^{K} \mathbf{H}_k \mathbf{Q}_k \mathbf{H}_k^\dagger$
    $\hat{\mathbf{H}}_i = \left(\mathbf{W}_z^{(i)}\right)^{-1/2} \mathbf{H}_i$
    $\mathbf{Q}_i = \underset{\operatorname{diag}(\mathbf{Q}) \preccurlyeq \mathbf{P}_i, \mathbf{Q} \succcurlyeq 0}{\arg\max} \log\det(\hat{\mathbf{H}}_i \mathbf{Q} \hat{\mathbf{H}}_i^\dagger + \mathbf{I}_m)$
    (using single-user mode-dropping in Section III.A)
  end
**until the sum rate converges**

---

The following theorem establishes the sure convergence of the proposed algorithm:

**Theorem 2** (Convergence). *In iterative mode-dropping for MIMO-MAC, the sum rate always converges to the sum capacity and all the covariance matrices $\{\mathbf{Q}_i\}$ converge to the optimal matrices.*

*Proof:* At the $i$th step of each iteration, single-user mode-dropping finds the optimal $\mathbf{Q}_i$ to maximize the sum rate while keeping the other covariance matrices unchanged. Since the objective function of this subproblem is continuous and strictly concave, the optimal solution is unique and the optimized covariance matrix either remains unaltered or changes to increase the sum rate. Henceforth, the sum rate is nondecreasing at each step. Furthermore, it is always bounded above for a finite power budget. Hence the sum rate must converge to a limit. In addition, the sum rate is also a continuous function of $\mathbf{Q}_i$. Therefore, as the sum rate approaches the sum capacity, the changes in $\mathbf{Q}_i$ diminish. In the end, each covariance matrix converges to the solution of the single-user problem in (7). According to Theorem 1, the proof is complete. ∎

*Remark* 1: Iterative water-filling and iterative mode-dropping update $\{\mathbf{Q}_i\}$ differently. At each step of iterative water-filling, the eigenvectors of $\mathbf{Q}_i$ are solely determined by the left singular vectors of corresponding effective channel matrix and do not depend on the SNR of the $i$th user. The eigenvalues of $\mathbf{Q}_i$ are then found by water-filling the total power across the spatial subchannels. In contrast, for iterative mode-dropping, both the eigenvectors and eigenvalues of the optimal $\mathbf{Q}_i$ depend not only on the effective channel matrix but also on the SNR and the power constraint. This is because with per-antenna power constraint, the optimal $\mathbf{Q}_i$ is not diagonalizable by the left singular vectors of the effective channel matrix. Thus in each step, iterative mode-dropping usually updates both the eigenvectors and eigenvalues of $\mathbf{Q}_i$ simultaneously.

*Remark* 2: The convergence in Theorem 2 holds for all random channel realizations. Although Theorem 2 holds for arbitrary initial conditions, setting initial covariance matrices to zero provides a fast convergence. More specifically, the sum rate is at most $(K-1)m/2$ nats away from the sum capacity after only one iteration for every $\mathbf{Q}_i$. This is the same as the rate of convergence for iterative water-filling. Since the proof follows the same procedure as the proof for Theorem 3 in [2], we only highlight here the difference. The terms $\lambda_i P_i$ and $\mathbf{S}_z$ in [2] are replaced by $\text{tr}(\mathbf{D}_i \mathbf{P}_i)$ and $\mathbf{I}_m$, respectively. To illustrate that the multiuser duality gap is bounded by $(K-1)m/2$, the authors of [2] proved that the dual variables $\lambda_i \leq \lambda'_i$. Here, instead, we need to show that $\text{tr}(\mathbf{D}_i \mathbf{P}_i) \preccurlyeq \text{tr}(\mathbf{D}'_i \mathbf{P}_i)$ for all $i = 1, 2, \cdots, K$, where diagonal matrix $\mathbf{D}_i$ is the solution of the following minimization problem:

$$\begin{aligned}\min \quad & \text{tr}(\mathbf{D}_i \mathbf{P}_i) \\ \text{s.t.} \quad & \mathbf{D}_i \succcurlyeq \mathbf{H}_i^\dagger \mathbf{W}_K^{-1} \mathbf{H}_i\ ,\end{aligned} \quad (11)$$

and diagonal matrix $\mathbf{D}'_i$ is the solution of the following minimization problem:

$$\begin{aligned}\min \quad & \text{tr}(\mathbf{D}'_i \mathbf{P}_i) \\ \text{s.t.} \quad & \mathbf{D}'_i \succcurlyeq \mathbf{H}_i^\dagger \mathbf{W}_i^{-1} \mathbf{H}_i\ ,\end{aligned} \quad (12)$$

where $\mathbf{W}_i = \sum_{k=1}^{i} \mathbf{H}_k \mathbf{Q}_k \mathbf{H}_k^\dagger + \mathbf{I}_m$. Indeed, because

$$\mathbf{H}_i^\dagger \mathbf{W}_K^{-1} \mathbf{H}_i \preccurlyeq \mathbf{H}_i^\dagger \mathbf{W}_i^{-1} \mathbf{H}_i, \quad \forall i = 1, 2, \cdots, K\ , \quad (13)$$

the feasible region of the optimization problem in (11) includes that of (12) as a subset. Therefore, at optimum, we have $\text{tr}(\mathbf{D}_i \mathbf{P_i}) \preccurlyeq \text{tr}(\mathbf{D}'_i \mathbf{P_i})$, which holds for all $i = 1, 2, \cdots, K$.

*Remark* 3: The computational complexity of iterative mode-dropping is approximately the product of the expected complexity of single-user mode-dropping, the number of iterations and the number of users $K$. We found from numerous numerical experiments that the former two almost do not change and therefore, the total complexity scales linearly with $K$ as shown later in simulations (Figure 2). In contrast, the complexity of interior-point methods is a cubic function of $K$ [6]. This improvement becomes more significant when the number of users $K$ increases.

*Remark* 4: Although the proposed algorithm successfully calculates the sum capacity of MIMO-MAC under per-antenna power constraint, it cannot be used directly to generate the whole capacity region except the inner and outer bounds as shown in Section IV. To obtain the whole region, one needs to solve the following weighted sum-rate maximization problem:

$$\begin{aligned}\max_{\{\mathbf{Q}_i\}} \quad & \sum_{j=1}^{K} \omega_{\pi_j} \log \det \left( \frac{\mathbf{I}_m + \sum_{i=1}^{j} \mathbf{H}_{\pi_i} \mathbf{Q}_{\pi_i} \mathbf{H}_{\pi_i}^\dagger}{\mathbf{I}_m + \sum_{i=1}^{j-1} \mathbf{H}_{\pi_i} \mathbf{Q}_{\pi_i} \mathbf{H}_{\pi_i}^\dagger} \right) \\ \text{s.t.} \quad & \text{diag}(\mathbf{Q}_i) \preccurlyeq \mathbf{P}_i, \quad i=1,\cdots,K \\ & \mathbf{Q}_i \succcurlyeq \mathbf{0}, \quad i=1,\cdots,K\ ,\end{aligned} \quad (14)$$

where $\pi_i$ denotes decoding order (the user $\pi_K$ is decoded first and user $\pi_1$ decoded last) and the weights satisfy $\omega_{\pi_1} \geq \omega_{\pi_2} \geq \cdots \geq \omega_{\pi_K}$. Efficient solution for this weighted-sum-rate problem up to now is still unclear, even under the sum power constraint.

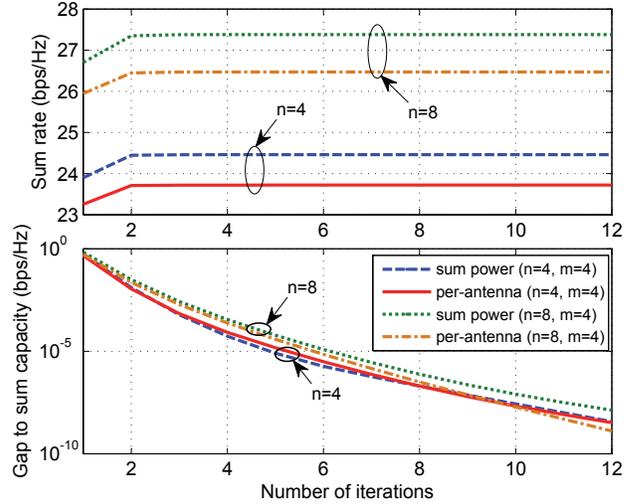

Fig. 1. Convergence examples of the sum rate under per-antenna power constraint (iterative mode-dropping) and sum power constraint (iterative water-filling).

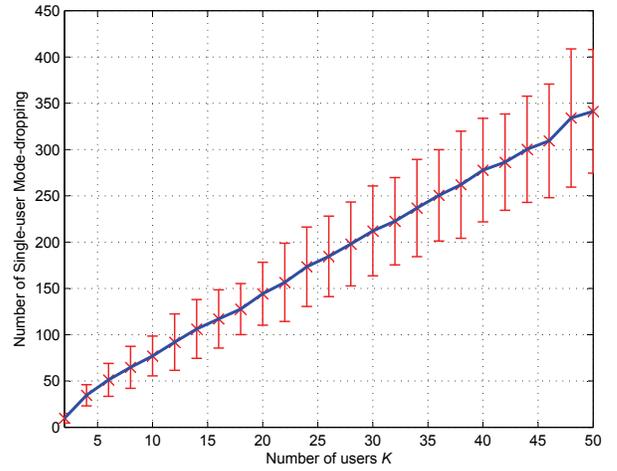

Fig. 2. The expected number of single-user mode-dropping iterations required to reach within $10^{-6}$ bps/Hz of sum capacity, together with the standard deviation.

## IV. NUMERICAL EXAMPLES

In this section, we numerically investigate the convergence behavior of iterative mode-dropping and quantify the difference between the MIMO-MAC capacities under the per-antenna and the sum power constraints. Throughout this section, we assume that the wireless channels undergo independent Rayleigh fading, that is, the entries of the channel matrices are i.i.d. circularly symmetric complex Gaussian variables with zero mean and unit variance. The ergodic capacity is computed by averaging over 2000 channel realizations.

In Figure 1, we compare convergence behaviors of iterative mode-dropping and iterative water-filling. Consider a system of $K=15$ users with two different antenna settings: $n=4, m=4$ and $n=8, m=4$. For each user, the per-antenna power constraint is set to 0.5 for $n=4$ and 0.25 for $n = 8$ and the sum power constraint is set to 2. Figure 1 shows the ergodic sum rate versus the number of iterations for both power constraints. After the first iteration, the sum rates are already very close to the sum capacity, and after the second iteration, the gap to optimality is lower than $10^{-1}$ for both constraints. The same convergence behavior can be observed for all simulations that we ran. In addition, the higher capacity achieved with more transmit antennas is attributed to the diversity gain and array gain. In Figure 2, the expected number of single-user mode-dropping iterations required to reach within $10^{-6}$ bps/Hz of the sum capacity versus the number of users is shown together with the corresponding standard deviation. Here we let $n = m = 4$, the per-antenna constraint be 0.5, and use 500 random channel realizations to generate these statistics. This result verifies Remark 3 that the total complexity grows linearly with the number of users $K$.

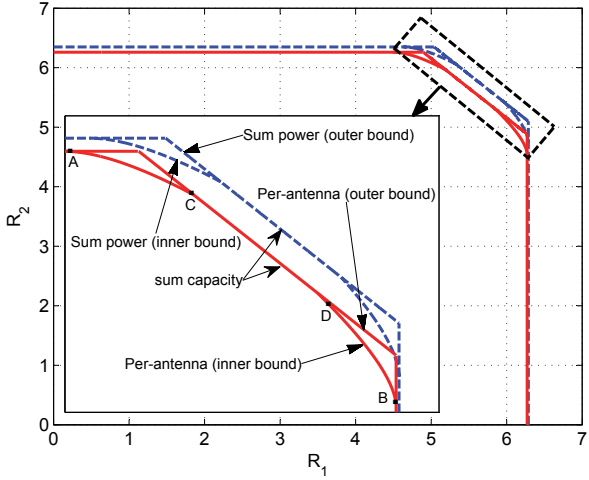

Fig. 3. Sum capacities and bounds on the capacity region of the two-user MIMO-MAC under per-antenna or sum power constraint.

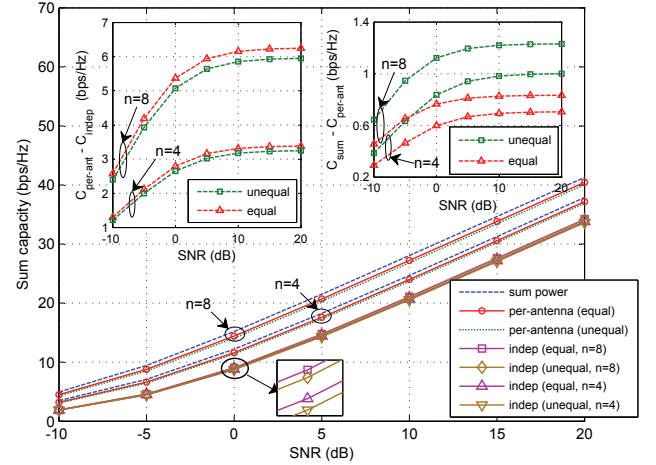

Fig. 4. Comparison of the ergodic sum capacity versus per-user SNR for different scenarios. The insetted figures show the capacity gaps between the per-antenna constraint and independent spatial multiplexing (left), and between the sum power and per-antenna power constraints (right). Parameters: $K = 4, m = 4$ and $n = 4, 8$.

Figure 3 presents the capacity region of a two-user MIMO-MAC for the two different power constraints. The numbers of antennas are $n = m = 4$, the sum power constraint is 2 per user and the per-antenna constraint is 0.5. The first iteration of all the transmit covariance matrices leads to $\{\mathbf{Q}_1^A, \mathbf{Q}_2^A\}$ that can achieve the point A if $\mathbf{Q}_2$ is updated before $\mathbf{Q}_1$, and $\{\mathbf{Q}_1^B, \mathbf{Q}_2^B\}$ that can achieve the point B vice versa. Iterative mode-dropping converges to both point C and point D, which give the sum capacity line CD. The inner-bound curve AC corresponds to rate region achieved by the convex hull $\{\mu \mathbf{Q}_1^A + (1-\mu)\mathbf{Q}_1^C, \mu \mathbf{Q}_2^A + (1-\mu)\mathbf{Q}_2^C\}$ with $0 \leq \mu \leq 1$, similar for the curve BD. The piecewise outer bound comprises of three line segments that are defined by the single-user capacities and the sum capacity. The proposed algorithm converges precisely to the sum capacity with per-antenna constraint and can provide computationally simple inner and outer bounds to the capacity regions. In addition, this figure illustrates the difference between the capacity regions under the two power constraints.

Figure 4 compares the sum capacities under the following scenarios: 1) sum power constraint; 2) per-antenna constraint with equal power across antennas; 3) per-antenna constraint with unequal power, in which the power of the $k$th antenna is proportional to $k$; 4) spatial multiplexing with equal substreams, that is, the transmit covariance matrices $\mathbf{Q}_i$'s are proportional to $\mathbf{I}$; and 5) spatial multiplexing with unequal substreams, in which $\mathbf{Q}_i$'s are diagonal, with the $k$th diagonal entry proportional to $k$. We also plot the capacity gaps between the sum power and the per-antenna power constraints (equal or unequal), and between the per-antenna constraints (equal or unequal) and the corresponding independent spatial multiplexing. The parameters are chosen as $K = 4$, $m = 4$, and $n = 4, 8$.

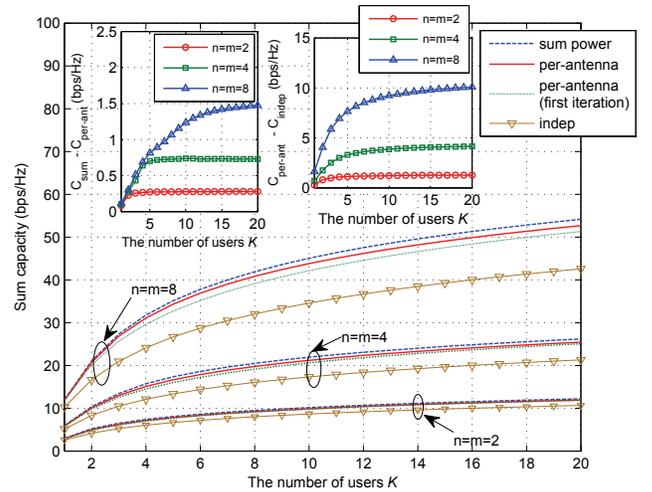

Fig. 5. Ergodic sum capacity versus the number of users under sum power constraint, per-antenna power constraint and independent spatial multiplexing. For the per-antenna power constraint, the sum rate after the first iteration is also included. The insetted figures show the capacity gaps between the sum power and per-antenna constraints (left), and between the per-antenna constraint and independent spatial multiplexing (right).

The following observations can be made: First, as expected, the decreasing order of capacity is under sum power, per-antenna (equal power), per-antenna (unequal power), independent spatial multiplexing (equal power) and independent spatial multiplexing (unequal power). Our simulations provide a quantitative example of the difference between MIMO-MAC capacities under per-antenna constraint or sum power constraint, and show that both are significantly higher than the sum rate under independent spatial multiplexing in which $\mathbf{Q}_i$'s are not at all optimized. Second, as the number of transmit antennas $n$ increases, the increase in sum capacity under any of the first three power constraints is significantly more than that under independent spatial multiplexing. Additional transmit antennas provide a better diversity gain, and iterative waterfilling and iterative mode-dropping both attempt to minimize the effects of multiple access interference, whereas independent spatial multiplexing does not. Hence the capacity gap (compared to independent spatial multiplexing) increases with the number of transmit antennas.

Last, as illustrated in Figure 5, the sum capacity under sum power constraint, per-antenna power constraint and independent spatial multiplexing increases with the number of users $K$. The capacity gaps also increase with $K$ and the number of antennas. As $K$ increases, however, the capacity gaps become saturated because the system is more interference-limited. As the number of antennas increases (from $n = m = 2$ to $n = m = 8$) the capacity increases due to the degree of freedom gain. Furthermore, after just one iteration, the sum rate obtained by iterative-mode dropping is already in good proximity with the true capacity.

## V. Conclusion

In this letter, we have proposed an iterative mode-dropping algorithm to compute the sum capacity of the MIMO-MAC with per-antenna power constraint. It successively optimizes each user's transmit covariance matrix until convergence. Similar to iterative waterfilling, iterative mode-dropping is fully distributive and has fast convergence. The algorithm is attractive for its simplicity, low complexity and practicality when compared to general convex programing methods.

Simulations results illustrate the difference in the sum capacity of MIMO-MAC under per-antenna or sum power constraints. Both of them are significantly higher than the sum rate under independent spatial multiplexing in which the input signals are non-optimized. This gain in sum capacity, however, saturates with the number of users as the multiple access interference increases even under per-antenna power constraint. Increasing the number of antennas can increase the sum capacity as well as the capacity gain by having better degree of freedom gain, diversity gain or array gain.